\documentclass{ws-procs10x7}

\begin{document}

\title{Diffractive physics: from the Tevatron to the LHC}

\author{Michele Gallinaro}

\address{(for the CDF and the D\O~collaborations) \\ The Rockefeller University, 
1230 York Avenue, Box 188 - New York, NY 10021\\
E-mail: michgall@fnal.gov}  

\twocolumn[\maketitle\abstract{
Measurements of soft and hard diffractive processes have been performed at the Tevatron $p\bar p$ collider 
during the past decade.
Diffractive events are studied by means of identification of one or more rapidity gaps and/or a leading antiproton.
Here, results are discussed within the Tevatron data and compared to those obtained at the HERA $ep$ collider. 
The traditional ``pomeron'' is described within the framework of QCD and
the issues discussed include pomeron structure, diffractive cross section factorization,
and universality of rapidity gap formation.
Exclusive dijet and low-mass state production in double-pomeron exchange processes, 
including predictions for Higgs production at the LHC from dijet measurements at the Tevatron.}]

\section{Introduction}
\vspace*{-0.1cm}
At the Fermilab Tevatron collider, proton-antiproton collisions have been used 
to study diffractive processes at high energies.
Two experiments, CDF and D\O, collected data in the 1990's at an energy of $\sqrt{s}$=~1.8~TeV 
and continue to do so in the first decade of this century with new upgraded detectors 
during the second phase of data-taking at $\sqrt{s}$=~1.96~TeV.
These two periods are usually referred to as Run~I and Run~II, respectively.
Diffractive processes at the Tevatron are being
studied by tagging events with either a rapidity gap or a leading hadron.

\vspace*{-0.3cm}
\section{Rapidity gap formation}
\vspace*{-0.1cm}
Rapidity gaps, which are regions devoid of particles, are an important element in the description of diffractive processes,
and arise from the fact that the pomeron does not radiate as it is exchanged between the two hadrons.
In non-diffractive (ND) events, rapidity gaps are exponentially suppressed as a function of gap size.
In contrast, in the case of diffractive processes,
the dependence of the cross section on gap size
remains approximately constant even for large gaps.

The study of soft
processes in conjunction with the presence of one or more rapidity gaps can shed light 
on gap formation and survival probability and, furthermore,
help understanding the mechanism responsible for the suppression of the diffractive cross section 
relative to Regge predictions\cite{renormalization}.
In order to disentangle the effect of gap formation from
the presence of additional color-exchange processes,
events with two gaps were selected and the ratio of two-gap to one-gap rates was measured at the Tevatron.
One-gap rates were measured from the inclusive single diffractive (SD) sample in events which contain a leading antiproton tag; 
two-gap events were selected in a sub-sample with an additional central rapidity gap (SDD).
Measurements show that the differential shapes of event rates 
$dN/d\Delta\eta$ agree with Regge theory, 
but the ratio of two-gap to one-gap rates is not as suppressed as the ratio of one-gap to no-gap rates.
The suppression factor due to the formation of the second gap at $\sqrt{s}=1800~{\rm GeV}$ is measured to be 
$R=0.246\pm0.001 {\rm (stat)} \pm0.042 {\rm (syst)}$\cite{cdf_sdd}.

A simple model has been proposed\cite{dino}, which extends the renormalization model to multiple gap formation. 
In such a model,
the gap probability term is normalized to unity and a color-matching factor $k$ is included for each gap, such that for $n$ gaps, 
a reduction factor of $k^n$ is expected. However, in order 
to test this prediction for multiple ($>2$) gap formation,
one must wait for the Large Hadron Collider (LHC).

A process with two rapidity gaps in the final state, similar to the one discussed earlier, is double pomeron exchange (DPE).
In DPE events, both the leading proton and the leading anti-proton survive the interaction and escape in the very forward region, while
a pomeron is emitted from each nucleon and a pomeron-pomeron collision occurs.
Thus, the DPE event topology is characterized by large rapidity gap regions on both sides of the interaction.
The ratio of the inclusive DPE (two gaps) to SD (one gap) cross sections has been measured at the Tevatron\cite{incl_dpe}.
The DPE/SD cross section ratio yields $R=0.194\pm 0.001 (\rm stat)\pm 0.012 (syst)$
and it confirms the result obtained in the soft SDD processes discussed earlier.

In conclusion,
the diffractive cross section can be factorized into two terms: 
1) a gap formation probability, and 
2) the total cross section at the sub-energy of the diffractive mass.

\vspace*{-0.2cm}
\section{Hard diffraction}
\vspace*{-0.1cm}
High$-p_T$ jets may emerge from diffractively produced high-mass states. 
Such processes are usually referred to as hard diffractive and may provide insight on the nature of the pomeron.
The understanding of the mechanism that relates soft and hard processes may provide information on the transition 
between perturbative and non-perturbative QCD, since
gap formation in diffractive processes is a non-perturbative effect, while hard partonic scattering can be 
described within perturbative QCD.
Hard diffractive processes may help understand the nature of the pomeron by deciphering its partonic structure.
The structure of the pomeron, in terms of its quark and gluon content, can be probed in hard scattering processes
by comparing SD to ND event rates.
Using rapidity gaps in the forward regions, the fraction of diffractive candidates was measured in data samples 
containing $W$ or $Z$ boson, $b$-quark, dijet, and $J/\psi$ events.
The measured ratios of SD to ND event rates are all of the order of $\sim 1\%$ at $\sqrt{s}$=~1.8~TeV.
From the fractions of diffractive events in different processes, it is possible to estimate the gluon and quark
fractions in the pomeron.
Indeed, while dijet production is sensitive to both the quark and gluon component of the pomeron, 
$W$ production probes mainly the quark component, since it occurs through $q\bar q\rightarrow W$ 
at leading order (LO) QCD calculations.
Although the processes studied have different sensitivities to quark and gluon content fractions in the pomeron, 
the measured fraction of diffractive events is approximately the same in all cases.
It therefore appears that the structure of the pomeron is not very different from the structure of the proton.
Combining the results from dijet, $W$-boson, and $b$-quark production, the gluon fraction 
in the pomeron was measured to be $f_g=0.54^{+0.16}_{-0.14}$~\cite{bquark}, 
in agreement with the measurement at HERA.

\vspace*{-0.2cm}
\section{Diffractive structure function}
\vspace*{-0.1cm}

\begin{figure}
\epsfxsize200pt
\figurebox{120pt}{160pt}{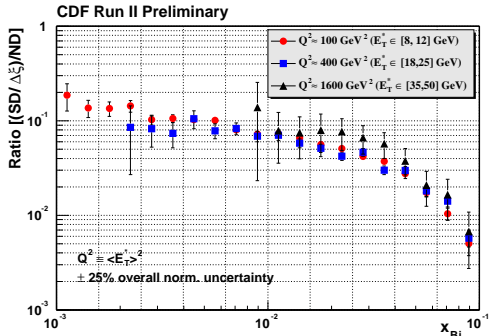}
\vspace*{-0.8cm}
\caption{Ratio of SD to ND dijet event rates.}
\label{fig:dsf_q2}
\end{figure}

Another interesting aspect to explore is the determination of the structure function of the pomeron, in order to understand how it 
relates to the structure of the proton.
Structure functions, i.e. the gluon and quark content of the interacting partons, 
can be investigated by comparing SD and ND events.
In LO QCD, the ratio of SD to ND dijet production rates is proportional 
to the ratio of the corresponding 
structure functions and can be studied as a function of the $x-$Bjorken scaling variable,
$x_{\bar p}=\beta\cdot\xi_{\bar p}$, of the struck parton in the antiproton, 
where $\beta$ is the pomeron momentum fraction carried by the parton. For each event, $x_{\bar p}$
is evaluated from the $E_T$ and $\eta$ of the jets using the equation
$ x_{\bar p}=\frac{1}{\sqrt s} \sum_{i=1}^nE_T^ie^{-\eta^i}$.
The ratio of SD to ND dijet production rates was measured at CDF\cite{dijet_sf} 
using the Roman Pot (RP) spectrometer to detect leading antiprotons.
Systematic uncertainties due to jet energy reconstruction and detector effects cancel out in the ratio.
The CDF result is suppressed by a factor of $\sim10$
relative to predictions from HERA data, indicating a breakdown of conventional factorization between HERA and the Tevatron. 
Using Run~II data, the CDF experiment measured SD to ND rates for dijet events with large transverse energy, where the jet energy spectrum extends to 
$E_T\sim 60~{\rm GeV/c^2}$. 
Preliminary results (Fig.~\ref{fig:dsf_q2}) indicate that the ratio does not strongly depend on $E_T^2\equiv Q^2$ 
in the range from $Q^2=100~{\rm GeV^2}$ to $1600~{\rm GeV^2}$, suggesting that the $Q^2$ evolution of the pomeron is similar to that of the proton.

In summary, 
three salient characteristics are noted in the measurement of the diffractive structure function: 
1) the SD structure function rises as $x_{\bar p}$ decreases, 
2) it has small, if any, $Q^2$ dependence, and 
3) the normalization of the SD to ND ratio is suppressed by a factor of $\sim 10$ with respect to the expectation of the 
parton distribution functions obtained at HERA. The latter implies that the pomeron does not possess a 
universal process-independent structure function, but the exchange is instead convoluted with additional effects, 
that spoil the formation of the rapidity gap.

\vspace*{-0.2cm}
\section{Exclusive production}
\vspace*{-0.1cm}

\begin{figure}
\epsfxsize200pt
\figurebox{120pt}{160pt}{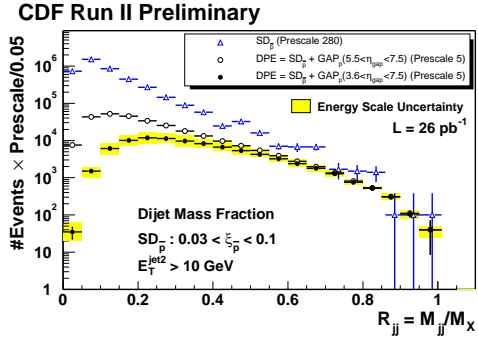}
\vspace*{-0.8cm}
\caption{Dijet mass fraction in DPE events.}
\label{fig:excl_dpe}
\end{figure}

The search for the Higgs boson occupies the central stage of the high-energy physics program both
currently at the Tevatron and in the near future at the LHC at CERN.
In the case of a small Higgs mass,
diffractive processes, with lower branching ratios,
may result in a cleaner experimental signature and are worth exploring.
This is the case of the Higgs production through DPE processes.
Diffractive production of the Higgs boson can be searched for at hadron colliders in the process
$p\bar p \rightarrow p H \bar p$ (or $p p \rightarrow p H p $),
where the leading hadrons in the final state are produced at small angles with respect to the direction of the incoming particles
and two large rapidity gap regions are present on opposite sides of the interaction.
In such events, the diffractive Higgs production could provide a distinct signature with  
exclusive two-jet ($b\bar b$ or $\tau^+\tau^-$) event final states
and two large rapidity gaps on both sides of the interaction.
The presence of the rapidity gaps provides an experimental environment which is practically
free of soft secondary particles and where the signal to background event ratio is favorable. 
In fact, the background from direct $b\bar b$ production is small thanks to several suppression mechanisms
(such as color and spin factors, and the $J_z=0$ selection rule).
Furthermore, the signal from $H\rightarrow b\bar b$ is expected to have a mass resolution which is
greatly improved due to the absence of secondary particles.
Predictions for the Higgs cross section due to exclusive DPE production are model dependent. In one of these models\cite{khoze_higgs},
which at the time of writing has still survived the exclusion limits set by the Tevatron data, and for a Higgs mass of $M_H=120~{\rm GeV/c^2}$, 
the predicted cross sections are $\sigma_H^{\rm TeV}\sim 0.2$~fb at the Tevatron
and $\sigma_H^{\rm LHC}\sim3$~fb at the LHC, with large uncertainties. 
Even according to these optimistic calculations, only a handful of events are expected for each $100 ~{\rm fb^{-1}}$ of data at the LHC, 
suggesting that this channel may be hard to unveil. 
In this context, the exclusive dijet production rate of the DPE events is of great interest in 
determining
the (background to) exclusive Higgs production cross section and to prepare for the LHC experiments. 
During Run~I, about 100 DPE candidate events were identified and used to set an upper limit on a
exclusive dijet production cross section\cite{dpe}.
At CDF, in $26~{\rm pb^{-1}}$ of Run~II data, the DPE final sample already consists of approximately 17,000 events.
The dijet mass fraction ($R_{jj}$), defined as the dijet invariant mass ($M_{jj}$) divided by the mass of the entire system, 
$M_X =\sqrt{\xi_{\overline{p}}\cdot\xi_p\cdot s}$, is calculated using all available energy in the calorimeter.
If jets are produced exclusively, $R_{jj}$ should be equal to unity.
In Figure~\ref{fig:excl_dpe} there is no visible excess evident over a smooth distribution.
After including systematic uncertainties, an upper limit on the exclusive dijet production cross section is calculated
based on all events with $R_{jj}>0.8$. 
The measurements provide an extremely generous upper limit cross section as all events at $R_{jj}>0.8$ 
are considered candidates for exclusive dijet production.
The cross section upper limit was 
set at $\sigma_{excl}^{TeV}< 1.0 (0.03)$~nb for jets with $E_T>10 (25) ~{\rm GeV}$.

A process similar to exclusive Higgs production is the exclusive production of $\chi^0_c$ (or $\chi^0_b$), 
as it has the same quantum numbers as the Higgs boson.
This process occurs in the 
$p \bar p\rightarrow p \chi^0 \bar p$ channel (where $\chi^0\rightarrow \mu^+\mu^-\gamma$)
and has been searched for at CDF during Run~II using 93~pb$^{-1}$ of data collected with a di-muon trigger.
Given the small number of events found in the final sample, it is experimentally difficult to 
evaluate the final contribution from background. 
Therefore, the 10 events found are to be considered as an upper limit on the exclusive production cross section of
$\sigma(p\bar p\rightarrow p+J/\Psi+\gamma+\bar p)=49\pm18{\rm (stat)}\pm39{\rm (syst)}$~pb.
In summary,
exclusive $\chi$ production has not yet been found at the Tevatron and
the cross section upper limit is comparable to predictions\cite{khoze_chi}.

\vspace*{-0.2cm}
\section{Conclusions}
\vspace*{-0.1cm}

The results obtained  during the past decade have led the way to the identification of
striking characteristics in diffraction. Moreover, they have significantly contributed to an understanding of
diffraction in terms of the underlying inclusive parton distribution functions.
The regularities found in the Tevatron data and the interpretations of
the measurements can be extrapolated to the LHC era.
At the LHC, the diffractive Higgs can be studied but not without challenges, as triggering and event acceptance 
will be difficult to implement and improve.
Still, future research at the Tevatron and at the LHC holds much promise for further understanding of diffractive processes.

\end{document}